\documentclass[conference,a4paper]{APSIPA2025}
\usepackage{float}
\usepackage{stfloats}
\usepackage{booktabs}
\usepackage{amssymb}
\usepackage{amsmath}
\usepackage{graphicx}
\usepackage{multirow}
\usepackage{threeparttable}
\usepackage[backend=biber,style=ieee,maxnames=3,doi=false]{biblatex}
\AtEveryBibitem{%
  \clearfield{issn}%
  \clearfield{url}%
  \clearfield{urldate}%
  \clearfield{note}%
  \clearfield{howpublished}%
}
\usepackage{subfig}
\usepackage{graphicx}

\usepackage{geometry}
\usepackage{fancyhdr}

\fancypagestyle{firststyle}{
  \fancyhf{}
  \fancyhead[C]{2025 Asia Pacific Signal and Information Processing Association Annual Summit and Conference (APSIPA ASC)}
}

\begin{document}

\title{Dynamic Fusion Multimodal Network for \\ SpeechWellness Detection}

\author{
\authorblockN{
Wenqiang Sun\authorrefmark{1},
Han Yin\authorrefmark{2},
Jisheng Bai\authorrefmark{3}\authorrefmark{4}, and
Jianfeng Chen\authorrefmark{1}
}

\authorblockA{
\authorrefmark{1}Northwestern Polytechnical University, Xi'an, China \\
\authorrefmark{2}School of Electrical Engineering, KAIST, Daejeon, Republic of Korea \\
\authorrefmark{3}LianFeng Acoustic Technologies Co., Ltd., Xi’an, China \\
\authorrefmark{4}Xi’an University of Posts \& Telecommunications, Xi’an, China \\
Emails: \\
sunwenqiang@mail.nwpu.edu.cn, hanyin@kaist.ac.kr, \\
baijs@xupt.edu.cn, chenjf@nwpu.edu.cn (corresponding author)
}
}

\maketitle
\thispagestyle{firststyle}
\pagestyle{fancy}
\pagestyle{empty}

\begin{abstract}
  Suicide is one of the leading causes of death among adolescents. 
  Previous suicide risk prediction studies have primarily focused on either textual or acoustic information in isolation, the integration of multimodal signals, such as speech and text, offers a more comprehensive understanding of an individual’s mental state. 
  Motivated by this, and in the context of the 1st SpeechWellness detection challenge, we explore a lightweight multi-branch multimodal system based on a dynamic fusion mechanism for speechwellness detection.
  To address the limitation of prior approaches that rely on time-domain waveforms for acoustic analysis, our system incorporates both time-domain and time-frequency (TF) domain acoustic features, as well as semantic representations.
  In addition, we introduce a dynamic fusion block to adaptively integrate information from different modalities. 
  Specifically, it applies learnable weights to each modality during the fusion process, enabling the model to adjust the contribution of each modality.
  To enhance computational efficiency, we design a lightweight structure by simplifying the original baseline model. 
  Experimental results demonstrate that the proposed system exhibits superior performance compared to the challenge baseline, achieving a 78\% reduction in model parameters and a 5\% improvement in accuracy.
\end{abstract}

\section{Introduction}
Suicide is one of the main causes of death among adolescents \cite{siucide_main_cause}. 
Timely identification of suicidal tendencies among adolescents and appropriate intervention are effective strategies for preventing adolescent suicide \cite{suicide_main_resson_death}. 
Suicide is affected by many factors, such as social psychology, culture, region and economy. 
Previous identifications of suicidal tendencies mainly rely on doctors' clinical observations, assessments, or patients' own expressions related to suicidal tendencies. 
However, these methods are time-consuming and laborious, heavily relying on doctors' rich experience.

Research on suicide risk detection using machine learning has made significant progress. These approaches have substantially improved detection efficiency, and enabled large-scale screening. 
For instance, Sanderson et al. utilized outpatient visit records, hospitalization data, and community pharmacy dispensing information to extract 808 predictors per individual \cite{M_L_3}. By employing the optimal gradient boosted trees model, they achieved promising predictive performance. However, the data processed by the above methods are mostly structured data such as hospital medical records. For unstructured data such as chat records, social media text or voice interactions, machine learning methods face many challenges \cite{structured_unstructured}. 
 
In recent years, deep learning has shown superior performance in handling unstructured data, including text, speech, and behavioral cues \cite{audio_vision_cues}.
Arowosegbe et al. investigated the use of natural language processing (NLP) for suicide detection and prevention, emphasizing that integrating structured and unstructured data yields more accurate results than using either type alone \cite{NLP_suicide}. 
Tadesse et al. utilized data from Reddit social media\footnote{\url{https://www.reddit.com}} and applied word embedding techniques to improve text representation, demonstrating the strength of deep learning in building effective models for suicide risk assessment \cite{DL_Social_Media_Forums}. 
However, existing research has predominantly focused on textual data, such as clinical notes, self-reported questionnaires, and social media posts, while studies involving speech signals remain limited and underexplored \cite{AI_review_suicide_detection}.

Speech-based suicide risk monitoring offers several advantages, such as cost-effective and enabling continuous, non-invasive assessments \cite{review_speech}. 
Prior studies have demonstrated that individuals with suicidal tendencies often exhibit reduced speech efficiency, flattened prosody, monotonic delivery, and an overall lack of vocal energy \cite{acoustic_cure}. 
In particular, spectral characteristics of speech can effectively reflect subtle psychomotor and emotional cues associated with suicidal ideation, such as variations in energy distribution, pitch, and harmonic content \cite{ozdas2004investigation}.
Moreover, acoustic features such as Mel-Frequency Cepstral Coefficients (MFCC) and Mel-spectrogram capture speech frequency content over time and have been linked to mental health conditions like depression. Kumar et al. showed that Mel-spectrogram can effectively distinguish between depressed and non-depressed individuals \cite{MFCC_Depression}.
Following this line, the 1st SpeechWellness detection challenge\footnote{\url{https://SpeechWellness.github.io}} was held with the goal of assessing suicide risk in adolescents through multimodal analysis of speech and text \cite{wu20251stspeechwellnesschallengedetecting}.
The challenge provided a unique dataset consisting of speech recordings from 600 Chinese teenagers aged 10 to 18, with 50\% of the participants identified as at risk of suicide based on psychological scale assessments. 
The official baseline extracts audio features using Wav2vec 2.0 \cite{wav2vec2.0} and textual features using BERT \cite{BERT}.
However, the Wav2vec 2.0 in the audio branch extracts features directly from raw waveform inputs and lacks explicit frequency domain information, which can be highly informative for detecting suicide-related speech characteristics \cite{mel_suicide_detect}. 
In addition, the total parameters of the challenge baseline system reach approximately 427 million, posing significant challenges for deployment on resource-constrained edge devices and limiting practical scalability.

To address the above issues, we propose a lightweight, dynamically fused multimodal system for SpeechWellness detection.
We firstly incorporate acoustic features in the time-frequency (TF) domain into the model, leading to a multi-branch structure.
Considering that different modalities may contribute unequally to the final prediction, we design a dynamic fusion mechanism to adaptively adjust the importance of each modality.
In addition, due to the high model complexity of the stacked Transformer \cite{vaswani2017attention} layers in Wav2vec 2.0 and BERT, we simplify these components for a more lightweight model. 
Experimental results demonstrate that our approach not only improves detection performance  but also significantly reduces model complexity. 
Moreover, we conduct extensive comparisons between unimodal and multimodal models, highlighting the effectiveness of multimodal information fusion for SpeechWellness detection.
\begin{figure*}[t]
    \centering
    \includegraphics[width=\textwidth]{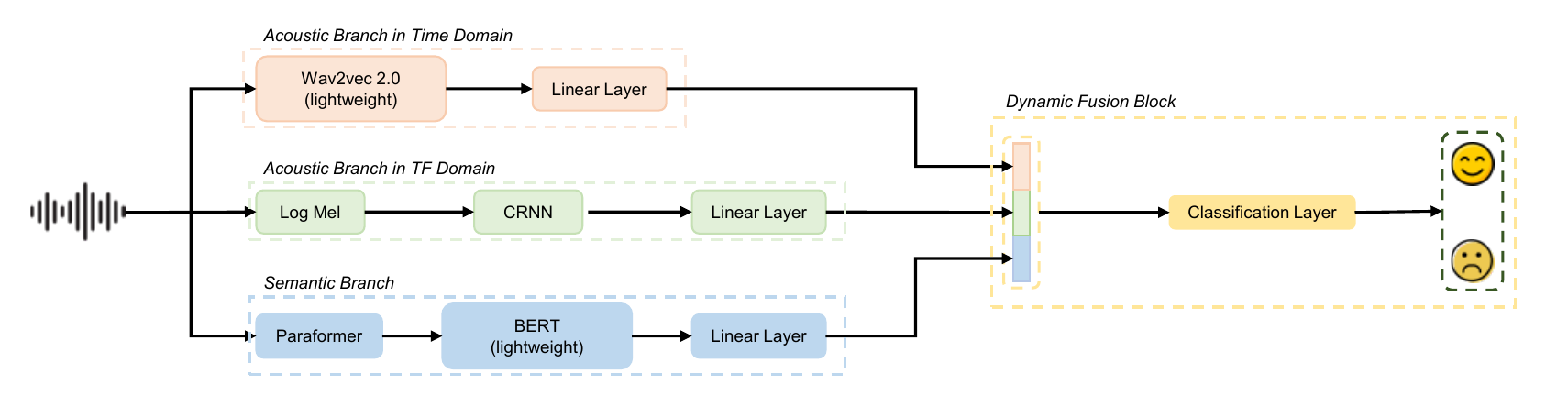} 
    \caption{Proposed dynamic fusion multimodal network for SpeechWellness detection}
    \label{fig1:overview}
\end{figure*}

\section{Methods}
As shown in Figure ~\ref{fig1:overview}, our model consists of three branches, i.e., acoustic branch in time domain, acoustic branch in TF domain, and semantic branch, which are used to process audio time domain features, audio TF domain features, and text features, respectively. 
We adaptively fuse the representations of the three branches through a dynamic fusion block.

\subsection{Acoustic Branch in Time Domain}
Deep learning models typically require large-scale labeled datasets to achieve satisfactory performance. However, collecting and annotating SpeechWellness detection datasets is time-consuming, especially given the ethical and privacy constraints involved \cite{wu20251stspeechwellnesschallengedetecting}. 
In scenarios with scarce labeled data, pre-trained models serve as a powerful foundation to enhance task-specific performance by providing generalizable representations \cite{wavlm}.

Wav2vec 2.0 \cite{wav2vec2.0} is a self-supervised speech representation model that learns high-level acoustic features directly from raw audio waveforms.
The model was pre-trained on 53k hours of speech, containing approximately 315 million parameters, achieving strong performance in downstream tasks with limited amounts of labeled data.
Therefore, we apply this pre-trained model as the time-domain acoustic feature extractor, followed by a linear layer for dimensional alignment.
However, due to the large number of parameters, deploying the full Wav2vec 2.0 model is relatively expensive and inefficient. 
 
In Wav2vec 2.0, the convolutional feature encoder captures low-level acoustic representations that are generally transferable across tasks, while the Transformer layers are primarily responsible for modeling contextual dependencies \cite{wav2vec2.0}. 
Prior studies have shown that reducing the number of Transformer layers has a relatively small impact on downstream task performance \cite{distill_Wav2vec}. 
Therefore, we significantly reduce the model size by retaining only the first four layers of the original 24-layer Transformer encoder, resulting in approximately 80\% parameter reduction.
We refer to this model as \textbf{Wav2vec 2.0 (lightweight)} in this work.

\subsection{Acoustic Branch in TF Domain}
\begin{figure}
    \centering
    \includegraphics[width=0.9\linewidth]{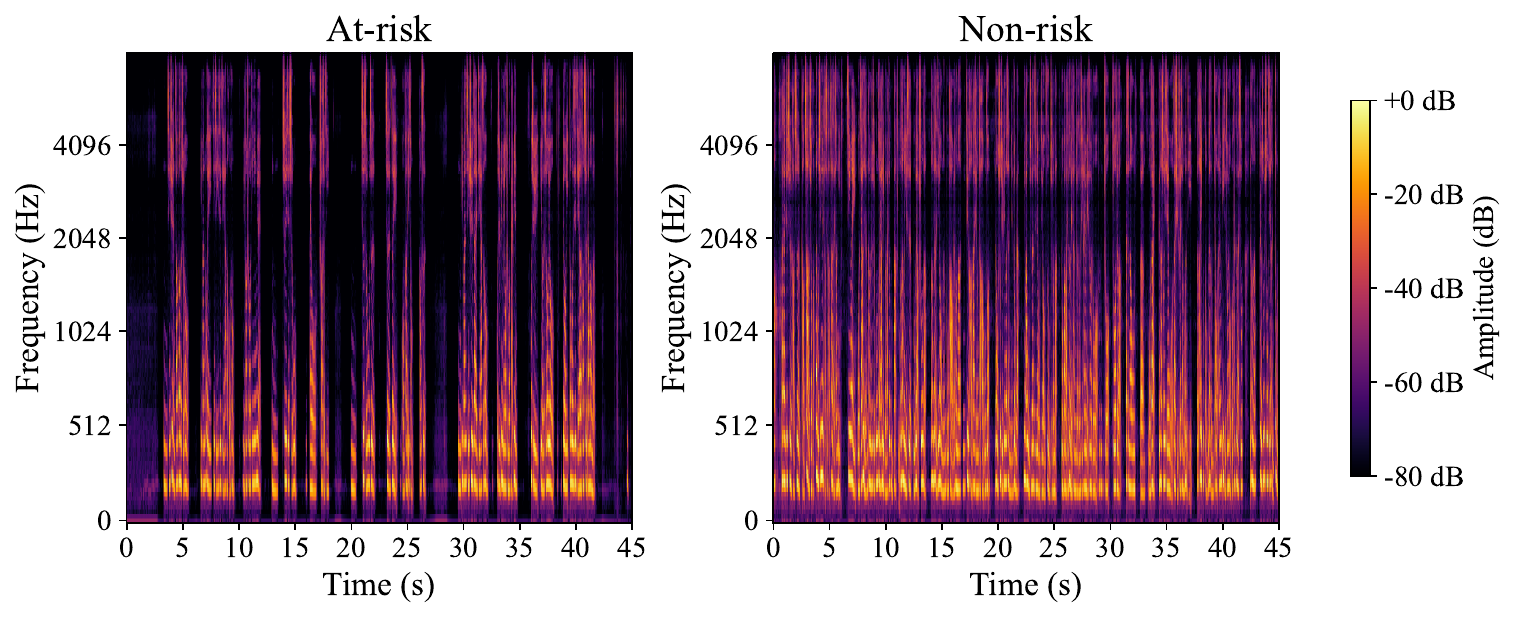}
    \caption{Mel-spectrogram comparison between suicidal risk (left) and non-suicidal risk (right) speech samples}
    \label{fig:mel_compare}
\end{figure}

Previous studies have demonstrated that frequency domain acoustic features are closely associated with mental health conditions such as suicidal tendencies \cite{freq_domain_ref}. These conditions are typically manifested in reduced pitch variability, lower energy, and flattened prosodic contours. 
For example, Figure ~\ref{fig:mel_compare} presents Mel-spectrograms of two speech samples. 
The speech sample from the suicidal individual (left) exhibits lower overall energy, reduced pitch variability, flatter prosodic contours, and more pronounced pauses compared to the non-suicidal sample (right). 
These are reflected by the weaker intensity across the Mel scale, less variation in frequency bands over time, and a more fragmented time-frequency structure.
Therefore, we incorporate an acoustic branch in TF domain to capture richer representations and further improve the detection performance.

TF domain acoustic features like Mel-spectrogram and MFCC are widely used in related research \cite{MFCC_used_widely}.
The Mel-spectrogram is computed by passing the audio signal through a bank of Mel-scale filters, producing features better aligned with human auditory perception. MFCC are derived by applying a Discrete Cosine Transform (DCT) to the Mel-filtered energies, yielding compact low-dimensional vectors.

\begin{figure}
    \centering
    \includegraphics[width=0.8 \linewidth]{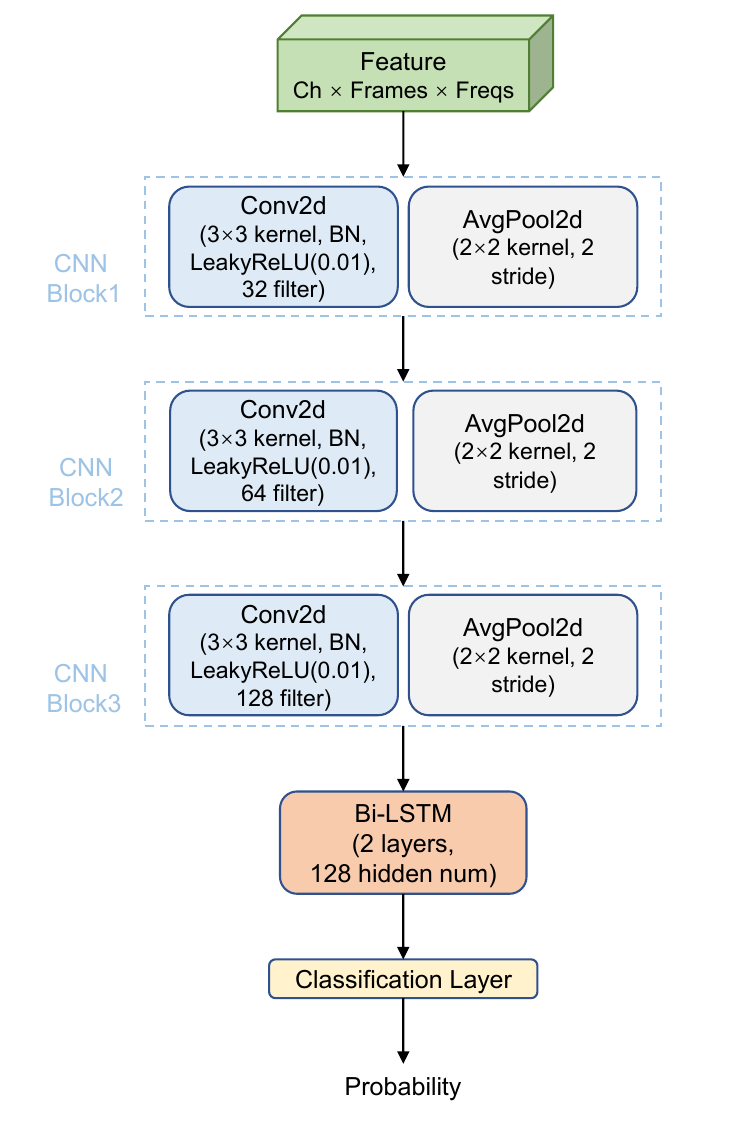}
    \caption{Architecture of the CRNN model used in acoustic branch in TF domain}
    \label{fig:crnn}
\end{figure}

Specifically, we use a Convolutional Recurrent Neural Network (CRNN) to extract acoustic representations in the TF domain. As shown in Figure ~\ref{fig:crnn}, the proposed CRNN model is composed of three two-dimensional convolutional (Conv2d) blocks, followed by a bidirectional Long Short-Term Memory (Bi-LSTM) \cite{LSTM} network and a fully connected classification layer.
The convolutional layers effectively extract local TF patterns from the spectrogram inputs, making them well-suited for modeling short-term spectral structures while significantly reducing the feature dimensionality.
The resulting representations are then passed to the Bi-LSTM, which models long-range temporal dependencies.
Finally, the hidden states from the forward and backward LSTM layers are concatenated and fed into a fully connected layer to generate the final prediction.
This architecture has demonstrated strong performance in various speech and audio classification tasks, while also offering the benefits of structural simplicity and ease of training \cite{AI_review_suicide_detection}.

\subsection{Semantic Branch}
Previous studies on suicide risk detection have primarily focused on textual content, including clinical records, questionnaires, EHRs, suicide notes, and posts from social media platforms.
These works, leveraging NLP, consistently demonstrate that textual information can provide strong cues for assessing suicide risk \cite{siucide_review_ML}. 
Following this line, we incorporate a semantic branch into the model for better performance.

Specifically, we first translate the speech into text with a state-of-the-art (SOTA) pre-trained automatic speech recognition model (i.e., paraformer \cite{paraformer}), then we employ a lightweight BERT as the text feature extractor. 
The BERT-base-Chinese model\footnote{\url{https://huggingface.co/google-bert/bert-base-chinese}}, which is based on the Transformer encoder architecture and was pre-trained on large-scale Chinese corpora, showing SOTA performance across various NLP tasks.
The model consists of two main components: an embedding layer and a Transformer encoder. 
The encoder comprises 12 Transformer blocks, each with 12 self-attention heads and a hidden size of 768, resulting in approximately 110 million parameters. 
This architecture enables the model to effectively capture deep contextual dependencies in Chinese text.

Similar to the strategy applied for the Wav2vec 2.0 (lightweight), we adopt a lightweight variant of BERT by retaining only the first Transformer block from the full encoder.  
This modification reduces the number of parameters by approximately 76\%, resulting in a total of 24.89 million parameters. 
Such a reduction makes the model more suitable for deployment in real-world applications without a substantial sacrifice in performance.
We refer to this model as \textbf{BERT (lightweight)} in this work.

\subsection{Dynamic Fusion Block}
Considering that the three branches in the network capture different aspects of suicide risk cues with varying levels of information richness and discriminative power. 
We apply a dynamic fusion block to adaptively fuse the multimodal information from the three branches.

Let $\mathbf{f}_t$, $\mathbf{f}_{tf}$, and $\mathbf{f}_s$ denote the feature vectors extracted from the time domain, TF domain, and semantic branches, respectively. Instead of directly concatenating these vectors, we assign each modality a learnable scalar weight: $w_t$, $w_{tf}$, and $w_s$. 
These parameters are initialized to 1 and optimized jointly with the model via backpropagation. The dynamic fusion strategy is formulated as:
\begin{equation}
\mathbf{f}_{\text{fused}} = \operatorname{concate}(w_t \cdot \mathbf{f}_t, \; w_{tf} \cdot \mathbf{f}_{tf}, \; w_s \cdot \mathbf{f}_s)
\end{equation}
where $\mathbf{f}_{\text{fused}}$ is the fused representation, which is then passed through a fully connected classification layer to obtain the final prediction, as formulated in:
\begin{equation}
\hat{y} = \text{softmax}(\mathbf{W} \mathbf{f}_{\text{fused}} + \mathbf{b})
\end{equation}
where $\mathbf{W}$ and $\mathbf{b}$ are the learnable weight matrix and bias vector of the classification layer.

This strategy allows the model to dynamically adjust the relative importance of each modality based on its contribution to the task, enhancing robustness in cases where one modality may be less or more beneficial.

\section{Experimental Setup}

\subsection{Dataset}
We use the dataset from the 1st SpeechWellness challenge for experiments, which contains speech recordings from 600 Chinese adolescents (420 females and 180 males) aged between 10 and 18 years, recruited from 47 elementary and middle schools in Guangdong Province, China. The participants were divided into 300 at-risk and 300 non-risk individuals based on clinical diagnoses using the  the Mini International Neuropsychiatric Interview for Children and Adolescents (MINI-KID) \cite{mini-kid} suicide assessment. 

In this dataset, each person was required to complete three tasks. 
Emotional Regulation (ER): an open-ended task in which subjects described their experiences of emotional distress and coping strategies. 
Passage Reading (PR): a structured task involving the reading of a standard paragraph (``The North Wind and the Sun'') to capture consistent linguistic features. 
Expression Description (ED): a semi-structured task where participants described the emotional content of a facial expression image. 
All recordings were conducted in Mandarin using standardized equipment in quiet, sound-proof settings to ensure audio quality and participant comfort.

The dataset is officially split into training (400 clips), validation (100 clips), and test (100 clips) sets.
The durations of the recordings for each task are presented in Table~\ref{tab:task_duration}.
Since the labels of the official test set are not publicly available, we use the validation set as a held-out test set for final evaluation.
Furthermore, we split the original training set into internal training and validation subsets (320:80) for model training and hyperparameter tuning.

\begin{table}[htbp]
\centering
\caption{Total duration (in hours) of speech recordings for each task and dataset split}
\label{tab:task_duration}
\resizebox{0.6\linewidth}{!}{
\begin{tabular}{lccc}
\toprule
\textbf{Split} & \textbf{Train} & \textbf{Dev} & \textbf{Test} \\
\midrule
Task0 (ER) & 3.17 & 0.93 & 0.96 \\
Task1 (PR) & 4.19 & 1.06 & 1.32 \\
Task2 (ED) & 2.10 & 0.55 & 0.63 \\
\midrule
\textbf{Total} & \textbf{9.45} & \textbf{2.55} & \textbf{2.92} \\
\bottomrule
\end{tabular}}
\end{table}

\subsection{Training Strategy}
In this work, the experimental setups mainly follow the challenge baseline.
Our training process consists of three stages. First, we fine-tune the acoustic branch in time domain using speech data. We add a classification layer after the linear layer for final prediction. 
Similarly, in the second stage, we fine-tune the semantic branch with the corresponding text content.
Finally, we freeze the fine-tuned acoustic branch in time domain and semantic branch, jointly training the acoustic branch in TF domain and the dynamic fusion block.
Following the challenge baseline, we train three systems for the three tasks, respectively. For inference, we average the outputs from the three systems as the final prediction.

\subsection{Detailed Configurations}
The Mel-spectrogram was computed with 128 Mel filter banks, using a window size of 40 ms and a hop size of 20 ms. The FFT size was set to 1024, and all audio samples were sampled at 16 kHz. For MFCC extraction, we used 40-dimensional coefficients with a frame length of 40 ms and a frame shift of 20 ms.
For model training, we apply the Adam optimizer with cross-entropy loss, all models are trained for a maximum of 200 epochs, with early stopping based on the validation loss.

In detail, follow the challenge baseline, in the first stage, we use a learning rate of $5\times10^{-5}$ for all the three tasks (i.e., ER, PR and ED), with a batch size of 8. In the second stage, the learning rates for the three tasks are $5\times10^{-5}$, $5\times10^{-5}$, and $5\times10^{-4}$, respectively, with a batch size of 16.
For the final stage, the learning rates are $1\times10^{-5}$, $3\times10^{-5}$, and $4\times10^{-5}$ for the three tasks, respectively, with a batch size of 8.


\section{Results and Discussions}
\subsection{Mel-spectrogram vs MFCC}
We first investigate the effectiveness of two commonly used TF features for SpeechWellness detection, including Mel-spectrogram and MFCC.
Specifically, we use a simple CRNN model (i.e., the acoustic branch in TF domain) for detection, with the two different TF features as the input.
Mel-spectrogram and MFCC are both computed based on the same 128 Mel filters to ensure comparability in frequency resolution.

The CRNN model is trained on each feature type separately, and results are shown in Table~\ref{tab:mel_mfcc_comp}. 
For each task, the classification accuracy is reported, along with the combined results, where the final prediction is made by averaging the softmax probabilities across the three tasks.

As shown in Table~\ref{tab:mel_mfcc_comp}, Mel-spectrogram shows slightly higher accuracy overall. 
Therefore, we use the Mel-spectrogram as the input feature of the acoustic branch in TF domain in the following experiments.

\begin{table}[ht]
\centering
\caption{Comparison of CRNN performance using MFCC and Mel-spectrogram features.}
\label{tab:mel_mfcc_comp}
\resizebox{0.8\linewidth}{!}{
\begin{tabular}{lcc}
\toprule
\textbf{Task} & \textbf{MFCC Acc. (\%)} & \textbf{Mel Acc. (\%)} \\
\midrule
Task 0 & 52.0 & 53.0 \\
Task 1 & 47.0 & 54.0 \\
Task 2 & 53.0 & 48.0 \\
Combined & 48.0 & 49.0 \\
\bottomrule
\end{tabular}}
\end{table}

\subsection{Monomodal vs Multimodal}
To explore the benefits of multimodal fusion, we evaluate the performance of individual branches, and the combination of the three branches. The results are summarized in Table~\ref{tab:modality_compare}.

\begin{table}[ht]
\centering
\caption{Comparison of mono- and multi-modal systems.}
\label{tab:modality_compare}
\resizebox{\linewidth}{!}{
\begin{tabular}{ccc|c}
\toprule
\textbf{Acoustic} & \textbf{Semantic} & \textbf{Feature} & \textbf{Acc. (\%)} \\
\midrule
Yes & No & Waveform & 51.0 \\
Yes & No & Mel & 49.0 \\
No & Yes & Text & 50.0 \\
Yes & Yes & Text \& Waveform & 52.0 \\
Yes & Yes & Text \& Mel & 53.0 \\
Yes & Yes & Text \& Mel \& Waveform \textbf{(proposed)} & \textbf{54.0} \\
\bottomrule
\end{tabular}}
\end{table}

The system without  pre-trained model shows the lowest performance, which is expected since pre-trained models have been exposed to large-scale datasets and are capable of capturing rich representations that are beneficial for SpeechWellness detection.
All multimodal systems outperform monomodal counterparts, showing the effectiveness of the multimodal feature fusion.
The combination of text and Mel-spectrogram modalities achieves an accuracy of 53\%, with an improvement of 3\% compared to the monomodal textual system.
While incorporating the waveform modality can bring an improvement of 2\%. 
This suggests that not all modalities contribute equally to the final detection, and demonstrates the necessity of adaptive fusion mechanisms, rather than naive feature concatenation.

\subsection{Comparison with Baselines}
\begin{figure}[htbp]
    \centering
    \subfloat[Proposed Model]{%
        \includegraphics[width=0.46\linewidth]{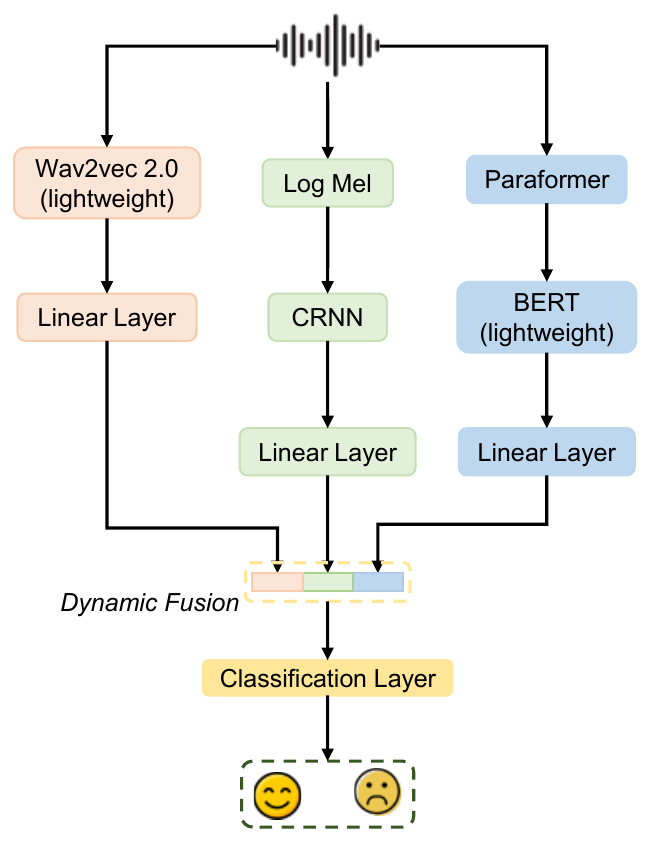}
    }
    \hfill
    \subfloat[Baseline Model]{%
        \includegraphics[width=0.45\linewidth]{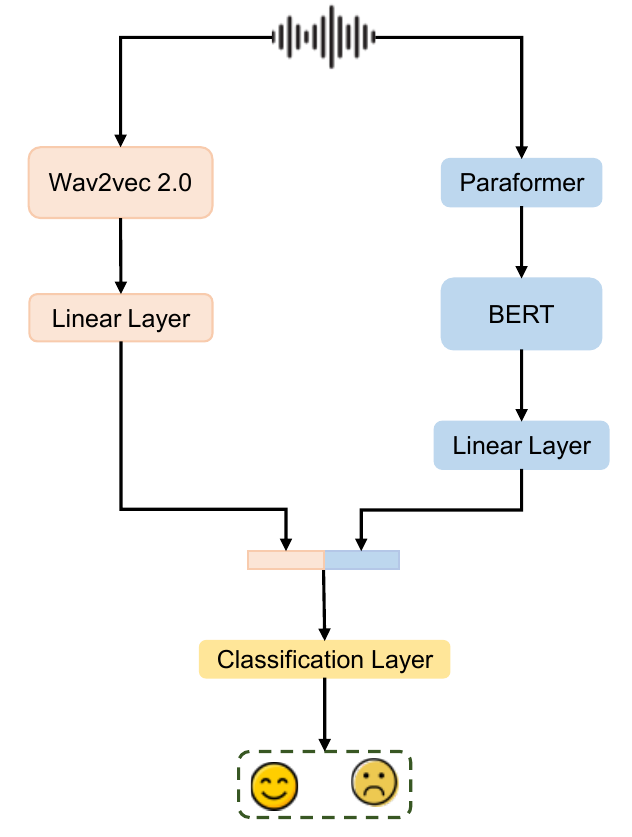}
    }
    \caption{Comparison between the proposed model and the baseline model.}
    \label{fig:models_comparison}
\end{figure}

The comparison between the proposed model and the baseline is shown in Figure ~\ref{fig:models_comparison}.
Baseline (lightweight) means we replace the Wav2vec 2.0 and BERT to the lightweight versions proposed in this work, i.e., Wav2vec 2.0 (lightweight) and BERT (lightweight).

\begin{figure}[htbp]
    \centering
    \includegraphics[width=\linewidth]{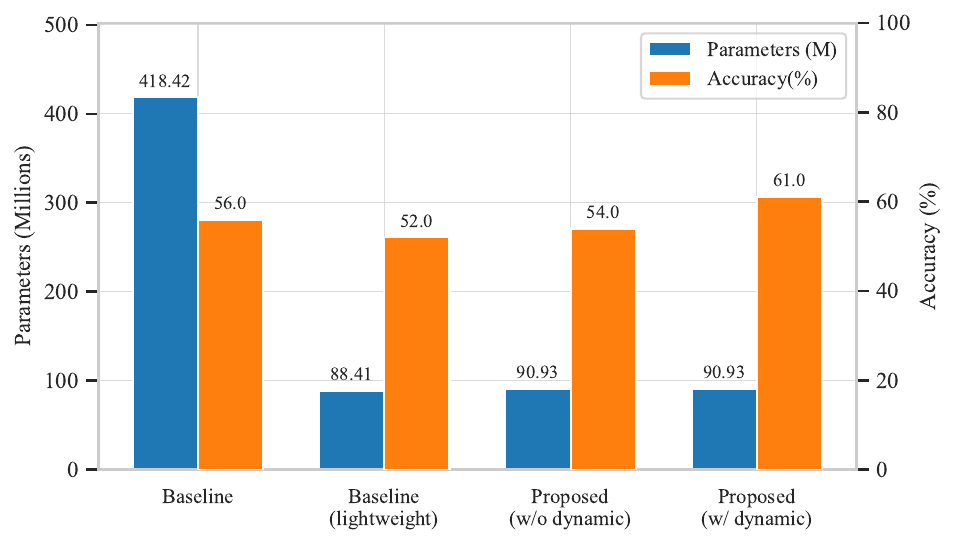}
    \caption{Results of the proposed and baseline systems (baseline, baseline-lightweight)}
    \label{fig:system compare}
\end{figure}

We compare our proposed system with the official baseline, as presented in Figure ~\ref{fig:system compare}.
First, the lightweight baseline results in an absolute accuracy drop of 4\% compared to the full-size baseline. 
By introducing an additional CRNN branch to extract TF domain features, the proposed model improves accuracy by 2\% over the lightweight baseline. 
Furthermore, incorporating the dynamic fusion mechanism leads to a 9\% absolute improvement in accuracy over the lightweight baseline. 
Overall, the proposed model achieves the best performance among all configurations while reducing the total parameter count to only 22\% of the original baseline, making it more efficient and suitable for edge deployment.

\section{Conclusions}
In this work, we propose a lightweight multi-branch multimodal network for SpeechWellness detection, which integrates time-domain, TF domain, and textual features through a dynamic fusion mechanism to more comprehensively capture indicators of suicide risk. 
First, we introduce acoustic features in the TF domain, enriching the representation beyond raw waveforms and enabling a multi-branch architecture. 
In addition, we design a dynamic fusion block to adaptively integrate multimodal information by weighting each modality according to its relevance.
Finally, we simplify the Wav2vec 2.0 and BERT to construct a more lightweight model to improve computational efficiency.
Experimental results show that multimodal systems outperform monomodal ones, by incorporating the TF domain acoustic feature and the dynamic fusion mechanism, the detection performance can be significantly improved.
Overall, the proposed model shows a 78\% reduction in model parameters and a 5\% accuracy improvement over the official challenge baseline. 
These results highlight the value of richer acoustic representations and efficient fusion strategies in speech-based mental health assessment.


\printbibliography
\end{document}